\documentclass[a4paper]{spicawStyle}
\usepackage{graphicx,natbib}

\newcommand{\OI}{O\,{\sc i}}
\newcommand{\CII}{C\,{\sc ii}}
\newcommand{\SI}{S\,{\sc i}}
\newcommand{\SiII}{Si\,{\sc ii}}
\newcommand{\NeII}{Ne\,{\sc ii}}
\newcommand{\FeII}{Fe\,{\sc ii}}
\newcommand{\emm}[1]{\ensuremath{#1}}   % Ensures math mode.
\newcommand{\emr}[1]{\emm{\mathrm{#1}}} % Uses math roman fonts.
\newcommand{\unit}[1]{\emm{\, \emr{#1}}}

\newcommand{\Lsun}  {\unit{L_\odot}}

\begin{document}

\title{Protoplanetary gas  disks in the far infrared}

\author{Javier R. Goicoechea\inst{1} and Bruce Swinyard\inst{2}\\ 
on behalf of the SPICA/SAFARI science team} 

\institute{Centro de Astrobiolog\'{\i}a, CSIC-INTA, Madrid, Spain
\and Rutherford Appleton Laboratory, Chilton, Didcot, Oxfordshire, UK}

\maketitle 

\begin{abstract}

The physical and chemical conditions in young protoplanetary disks set the boundary conditions 
for planet formation.   Although the dust in disks is relatively easily detected 
as a far-IR photometric ``excess'' over the expected photospheric emission, much less
is known about the gas phase. It seems clear that an abrupt \textit{transition} from massive 
optically
thick disks (gas--rich structures where only $\sim$1\% of the total mass is in the form of dust) 
to tenuous debris disks almost devoid of gas  occurs at $\sim$10$^7$ years, by which time the 
majority of at least the giant planets must have formed. Indeed, these planets are largely 
gaseous and thus they must assemble before the gas disk dissipates. Spectroscopic 
studies of the disk gas content at different evolutive stages are thus critical. 
Far-IR water vapor lines and atomic fine structure lines from abundant 
gas reservoirs (\textit{e.g.,}~[\OI]63$\mu$m, [\SI]56$\mu$m, [\SiII]34$\mu$m) are robust tracers 
of the  gas in disks. Spectrometers on board \textit{Herschel} will detect some of these lines 
toward the closest,  youngest and more massive protoplanetary disks. 
However, according to models, \textit{Herschel} will not reach the 
required sensitivity  to (1) detect the gas  residual in more evolved and 
tenuous \textit{transational disks} 
that are  potentially forming planets and  (2)  detect the gas emission from less massive 
protoplanetary disks around the most numerous stars in the Galaxy 
(M-type  and cooler dwarfs). Both are unique goals for SPICA/SAFARI. Besides,
SAFARI will be able to detect the far--IR modes of water ice at $\sim$44 and $\sim$62\,$\mu$m,
and thus allow water ice to be observed in many protoplanetary systems and fully explore its impact
on planetary formation and evolution.

\end{abstract}

\section{Introduction}

All planets are thought to form in the accretion disks that develop during the collapse and infall
 of massive dusty and molecular cocoons ($\geq$10,000~AU) where stars are born.  
However, we still have a very incomplete understanding of the physical and chemical conditions in such
circumstellar disks, how they evolve when dusty bodies grow and collide, which is their mineral content, how they 
clear as a function of time and, ultimately, how planets as diverse as the Earth or hot-Jupiters form around
different types of stars and at different places of the galaxy.

Circumstellar disks are divided into 3 classes according to their evolutionary stage. 
\textbf{Primordial protoplanetary disks}; which are rich in atomic and molecular gas, and are composed of relatively 
unprocessed interstellar material left over from the star formation process. Such young disks are very optically
thick in dust, with high radial midplane optical depths in the visual ($\tau_V>>1$).  
These protoplanetary disks start to become optically thin ($\tau_V\simeq1$) in a few million years after 
formation \citep{hais01} and evolve into \textbf{transitional disks} as their inner regions begin 
to clear at $\geq$10~Myr.  This is the critical intermediate stage when planetary formation is believed to take place, 
with dust particles colliding and growing to form larger bodies reducing the disk opacity. 
\textit{Spitzer} has shown that their outer regions (beyond $\sim$10--20~AU) can remain intact for longer, and thus 
residual gas can exist in the disk and play an important role in its evolution.  In spite of its importance, 
the residual gas content in \textit{transitional} disks at the first stages of planet formation 
is very poorly constrained, 
and many clues on these early stages of planetary formation can be provided by spectral line observations. 
 Disks with ages above $\geq$10~Myr are thought to be practically devoid of gas \citep{duve00} and the dust 
in these older debris disks is generally not primordial but continuously generated ``debris" from planetesimal 
and rocky body collisions.  The smallest dust grains have, at this stage, either been dispersed or have coagulated
into larger grains and the disk becomes very optically thin ($\tau_V << 1$). 
\textbf{Debris disks} are thus younger and more massive analogs of our own asteroid 
(\textit{hot inner disk}, $T_d\simeq$ 200~K) and Kuiper belts (\textit{cool outer disk}, $T_d\simeq$~60 K) so their 
study is vital to place the Solar System in a broader context (see 
more detailed contributions in these proceedings). 

%%--------------------------------------------------------------------------------------
\begin{figure*}[ht]
  \begin{center}
    \includegraphics[width=18.3 cm]{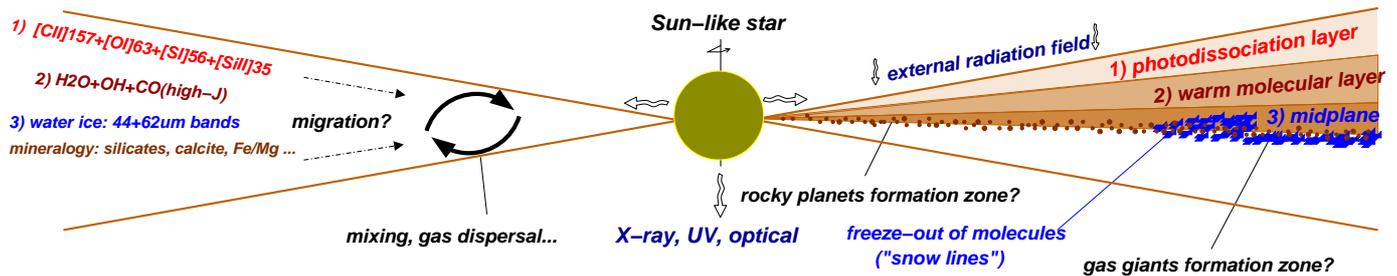}
  \end{center}
  \caption{Diagram showing where radiation arises from  a 
transitional protoplanetary disk (at the time when planets
are thought to assemble) and why the far-IR (i.e., SPICA/SAFARI) and the millimeter
 (e.g., \textit{ALMA}) together are essential to understanding the full picture of 
planetary formation and the primordial chemistry that may eventually lead
 to the emergence of life.}
\label{goicoecheaj1_fig:flared_disk}
\end{figure*}
%%-------------------------------------------------------------------------------------

\section{The gas content in planet forming disks}
\label{goicoecheaj1__sec:tit}

The physical and chemical conditions in young protoplanetary disks (\textit{primordial} and \textit{transitional}) 
set the boundary conditions for planet formation and an understanding of the formation and evolution of 
such disks will finally link star formation and planetary science. Planets are believed to form within the 
inner $\sim$20~AU of \textit{transitional} protoplanetary disks which emit most of their energy in the mid- and far-IR region 
\citep{boss08}. Although the dust is relatively easily  detected by photometric observations in the far-IR range, 
very little is known about the gas phase.  It seems clear that an abrupt transition from massive optically 
thick disks to tenuous debris disks occurs at $\geq$10~Myr \citep{meye08}, 
by which time the majority of at least the giant gaseous planets (e.g., ``Jupiters" and ``hot Jupiters")
 must have formed.  The very fact that these planets are largely gaseous means they must be formed before the 
gaseous disk dissipated, making the study of the gas in  protoplanetary disks essential to an understanding 
of how and where they formed.  For instance, the recently detected massive ``hot Jupiters" orbiting close to 
the parent star are very unlikely to have formed in situ, but rather they must have migrated inwards from outer 
disk regions beyond the \textit{snow line} (see Sect.~\ref{goicoecheaj1_sub-sec:ice}).  
Likewise it seems clear that Neptune and Uranus must have formed in a 
region closer to the Sun and migrated outwards.  
The mechanism for either of these scenarios is not certain.\\\\

Protoplanetary disk models predict a ``flared" structure (see Fig.~\ref{goicoecheaj1_fig:flared_disk}) which allows 
the disk to capture a significant portion of the stellar UV and X-ray radiation even at large radii 
\citep{qi06,ber07}
boosting the dust thermal emission, and the emission of mid- and far-IR lines from gas phase 
ions, atoms and molecules.  
Theoretical models recognise the importance of these X-UV irradiated surface layers, 
which support active photochemistry and are responsible for most of the mid- and far-IR line emission.  
They are, in many ways, similar to the well studied photodissociation regions of the ISM, 
that show a very rich far-IR spectrum (\textit{e.g.,} Goicoechea et al. 2004; 2009).   
Fine structure lines of the most abundant elements and metals (O, C, S, Si...) together with the
 far-IR rotational line emission of light hydrides (H$_2$O, OH...) are predicted to be the brightest gas 
cooling lines  of the warm disk, especially close to the star 
(Gorti \& Hollenach 2004, 2008; Woitke et al. 2009a, 2009b; Cernicharo et al. 2009). 

Indeed, far-IR [\SiII]34,  [\OI]63,145, [\SI]56 and [\CII]158 $\mu$m fine structure lines can be brighter 
than the mid--IR H$_2$ emission, the most abundant gas species in the disk. By detecting these 
lines we can  directly probe the physical and chemical conditions in disks around a large number of stars at differently
 evolutionary stages.  
\textit{Herschel} can only search for the strongest lines 
(above a few 10$^{-18}$\,W\,m$^{-2}$) toward the brightest, closest and
 most massive young protoplanetary disks.  In fact, only a few disks will be fully surveyed in the 
far-IR \textit{i.e.,} those with strong dust continuum which unfortunately difficulties 
the line detection  at low spectral resolution.  In particular \textit{Herschel} does not reach the sensitivity to 
either detect the  gas emission in the much more tenuous \textit{transitional} disks or detect the predicted gas line 
emission of less massive, and thus more numerous, disks around cooler stars.

%%----------------------------------------------------------------------------------------------
\begin{figure*}[ht]
  \begin{center}
    \includegraphics[width=15.7 cm]{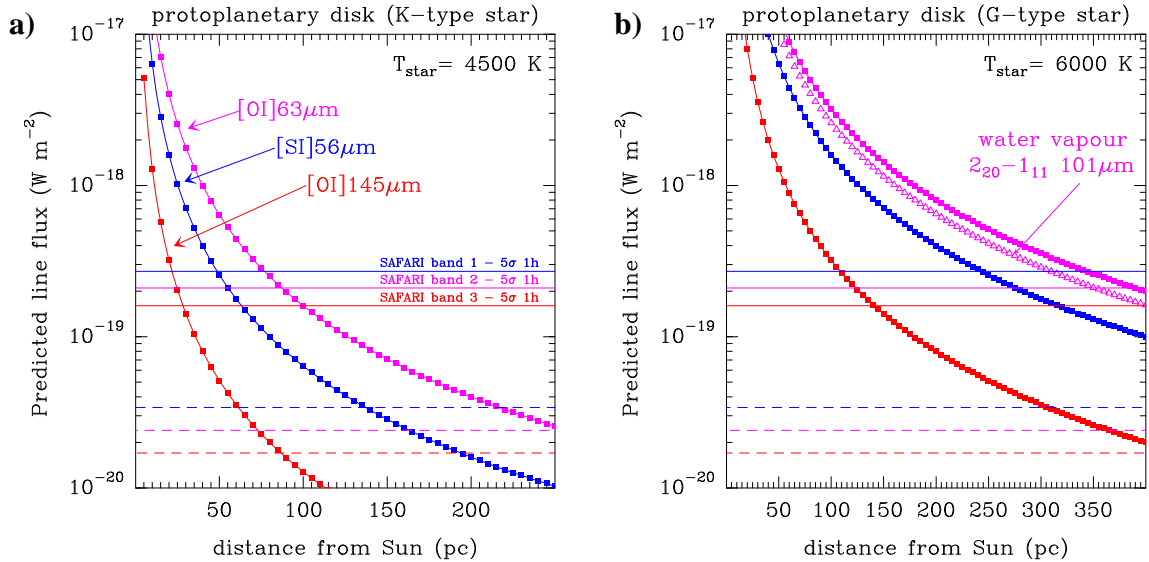}
  \end{center}
  \caption{Predicted fluxes of some key far-IR cooling lines in \textit{transitional}
protoplanetary  disks ($\sim$10Myr)  around \textbf{(a)} a cool K star 
 and \textbf{(b)} a G solar-type   as a function of distance. 
The horizontal lines indicate  5$\sigma$-1hr detection limits 
of SAFARI. 
Dashed  lines correspond to  sensitivities using narrow band filters on a restricted
number of wavelengths. 
The gas mass in the modelled disk is 10$^{-2}$\,M$_{Jup}$. %and the dust mass is 10$^{-5}$\,M$_{Jup}$. 
SPICA will have the sensitivity to observe
the gas content of disks in the closest star forming regions ($\sim$150\,pc) at the time 
planets assamble. Adapted from \cite{gort04} models.}
\label{goicoecheaj1_fig:gas_predictions}
\end{figure*}
%%---------------------------------------------------------------------------------------------

\section{SPICA studies of the gas dispersal}
\label{goicoecheaj1_sec-gas_dispersal}

In order to shed some light on the gas dispersal time scales, and thus on the formation of gaseous 
Jovian-type planets, high sensitivity infrared to submm spectroscopic observations over large statistical 
stellar  samples tracking all relevant disk evolutionary stages and stellar types are clearly needed.  
To date, studies of the gas content are biased to young and massive protoplanetary disks (probably not the 
most representative) through observations with ground-based (sub)mm interferometers and optical/near-IR 
telescopes.  Submm observations allows one to trace the outer and cooler disk extending over a few hundred 
AU \citep{dutr07}.  However, in this region most  molecular species freeze-out onto dust grain icy mantles 
and thus the determination of the total gas mass can be uncertain.  Recently the very inner regions of 
protoplanetary disks have been probed by means of optical/IR observations \citep{naji07}. 
 \textit{Spitzer} and ground-based telescopes have just started to show the potential diagnostic power of 
mid- and far-IR spectroscopy in a few  ``template" disks. This has allowed the exploration of the gas content and 
composition at intermediate radii (1-30~AU), \textit{i.e.}, the crucial region for the formation of planets.  
Surprisingly, recent mid-IR detections toward young disks include atomic lines such as 
[\NeII], [\FeII], etc. \citep{lahu07}, molecular species like H$_2$, H$_2$O, OH, HCN, C$_2$H$_2$ and  CO$_2$
\citep{carr08,saly08}, and even more complex organic species such as PAHs \citep{geer06,haba06} that
chemical models are now trying to reproduce (\textit{e.g.,} Ag\'undez et al. 2008).  
SPICA spectrometers will provide continuous wavelength 
coverage from $\sim$5 to 210~$\mu$m for the first 
time since the launch of ESA's \textit{Infrared Space Observatory} (1995).

Molecular hydrogen (H$_2$) is the most abundant gas species in a primordial protoplanetary disk 
($\sim$90\% of the initial mass).   Due to the relatively poor line sensitivity that can be achieved with 
even the largest ground-based mid-IR telescopes, H$_2$ has been detected only 
toward a few protoplanetary disks so far \citep{bitn08}. JWST will improve this situation enormously 
and many more disks will have been observed by the time SPICA is active.  
To complete and complement the JWST capability, SPICA's mid-IR high resolution echelle spectrometer (MIRHES) 
will be able to detect the brightest H$_2$ line (the $v$=0-0 $S$(1) line at $\sim$17~$\mu$m) 
with $\sim$10~km\,s$^{-1}$ resolution 
and much higher sensitivities than those achieved from the ground and with an order of magnitude higher spectral 
resolution than JWST.  Such a high spectral resolution will resolve the gas Keplerian rotation, trace the 
warm gas in the inner ($<$ 30~AU) disk regions and be sensitive to H$_2$ masses of the order of  
$\sim$0.1~M$_{Earth}$.

However, given the peculiar excitation conditions of H$_2$ (\textit{e.g.,} in the warm
UV illuminated edge of the disk  and in the hottest inner regions), 
the H$_2$ emission may not sample the whole circumstellar disk.  Instead, the deuterated isotopologue of molecular 
hydrogen, HD, although a few thousand times less abundant, can serve as a proxy of H$_2$ column density.  
The D/H abundance ratio of different  species can also be used as a diagnostic of the ISM origin 
of the young disk material.  HD has a small permanent dipole moment and its lowest energy rotational lines 
appear in the range  covered by SAFARI (112, 56 and 37~$\mu$m).

%%-------------------------------------------------------------------------------------
\begin{figure*}[ht]
  \begin{center}
    \includegraphics[width=16.7 cm]{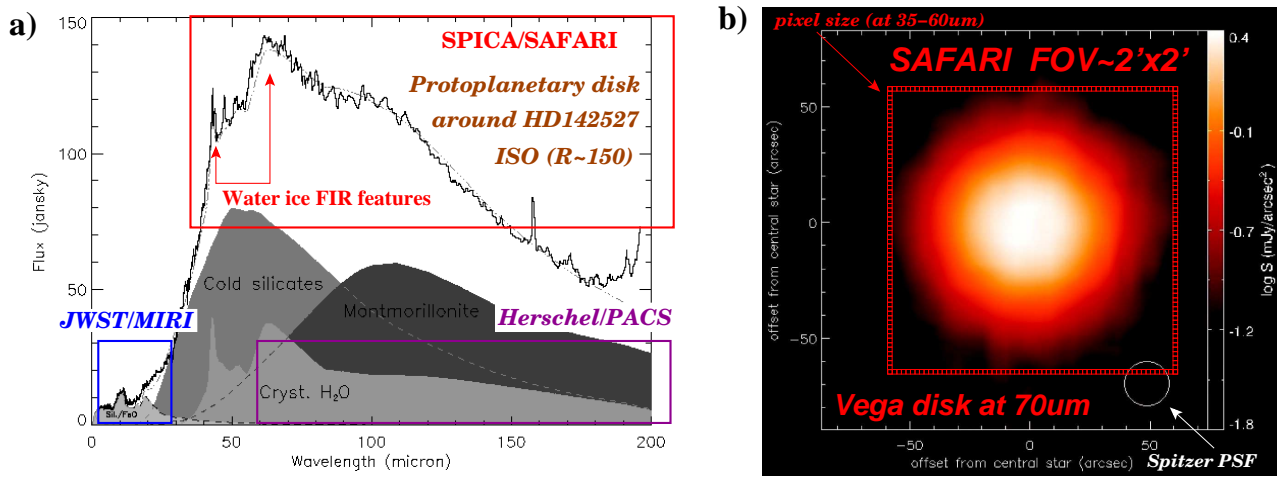}
  \end{center}
  \caption{\textbf{(a)} ISO spectrum of the disk around HD142527 \citep{malf99}. 
A these temperatures water ice can only be securely detected  in the far-IR.
Note that amorphous water ice only shows far-IR bands at $\sim$44\,$\mu$m \citep{moor92}.
Such features can not be observed  with \textit{Herschel} or JWST. 
SPICA will take the equivalent spectra of objects at flux levels less than 
10\,mJy per minute. \textbf{(c)} Image of  Vega debris disk at 70$\mu$m with
 \textit{Spitzer} \citep{su05}. SAFARI's large, fully sampled FOV and  
smaller pixel size  ($\sim$1.8$''$ at $\sim$44$\mu$m, red squares) 
 will provide very detailed spectro-images of nearby  disks.}
\label{goicoecheaj1_fig-disk-malfait}
\end{figure*}
%%------------------------------------------------------------------------

\subsection{Line detectability with SPICA/SAFARI}

At a distance of $\sim$50\,pc from the Sun, a given gas line 
will produce a flux of $\sim10^{-19}$(L$_{line}$/10$^{-8}$~\Lsun)\,W\,m$^{-2}$ where L$_{line}$ is the 
radiated power in the line expressed in solar luminosities.   SAFARI will be able to detect gas lines 
with luminosities $\geq10^{-8}~\Lsun$ in nearby, isolated, disks, or $\geq$10$^{-7}~\Lsun$ in more distant 
disks in the closest star forming regions at $\sim$150~pc (see Figure \ref{goicoecheaj1_fig:gas_predictions}).  
State-of-the-art  models of \textit{transitional} disks around Sun-like stars 
and with  gas masses as low as  $\sim$0.001~M$_{Jup}$ ($\sim$25~M$_{moon}$)  predict luminosities around 
10$^{-8}$~\Lsun\, for [\OI]145$\mu$m, [\SI]56$\mu$m and H$_2$O, OH or CO high-$J$ lines \citep{gort04}.
The [\OI]63$\mu$m line is expected to be the brightest line (10$^{-6}$\textendash 10$^{-7}$~\Lsun) even for 
modest gas masses and therefore it is one of the most robust gas tracers.  
The exciting prospect of detecting this small amount of gas in a statistically significant sample of disks,
\textit{e.g.,} toward the 6 closest ($\leq$150~pc) young stellar clusters with ages of $\simeq1-30$\,Myr: 
\textit{Taurus}, \textit{Upper Sco}, \textit{TW Hya}, 
\textit{Tuc Hor}, \textit{Beta Pic} and \textit{Eta Cha}, 
will provide a crucial test on the lifetime of gas  and  its dissipation timescales in planet 
forming disks  and therefore on the exoplanet formation theory itself.

\subsection{Chemical complexity in disks:}

In terms of chemistry, the protoplanetary disk is the major reservoir 
of key species with prebiotical relevance, 
such as oxygen, ammonia (NH$_3$), methane (CH$_4$)  or water (H$_2$O)  to be found later in planets, asteroids and 
comets. But how the presence and distribution of these species relates to the formation of planets, and most 
particularly, rocky planets with substantial amounts of water present within the so called 
\textit{habitable zone}, remains 
open to speculation without a substantial increase in observational evidence.   
Water is an obvious ingredient for life, 
and thus it is very important that we understand how water transfers from protostellar clouds and primordial 
protoplanetary disks to more evolved asteroids, comets and planets like our own.  Ultimately one has to understand 
how the water we see today in our oceans was delivered to the Earth. The ability of SAFARI to detect the emission 
of the critical species illustrated above will produce a unique meaningful census of disk chemical composition 
at different evolutionary stages and shed crucial insight on the chemical evolution of the gas before and during 
the process of planetary formation.  This in turn will tightly constrain the mechanisms by which water can be 
present in large quantities in the inner parts of planetary systems, \textit{i.e.,} the \textit{habitable zones}.

\subsection{Water ice in disks:}
\label{goicoecheaj1_sub-sec:ice}

Below gas temperatures of $\sim$150~K water vapour freezes-out onto dust grains and the main form 
of water in the cold circumstellar disk midplane and at large disk radii will be ice. 
 The physical location of the point at which water freezes out determines the position 
of the so called  ``\textit{snow line}", i.e. the water ice sublimation front, which separates 
the inner disk region of terrestrial, rocky, planet formation from that of the outer 
giant planets (Nagasawa et al. 2007; Woitke et al. 2009a, 2009b). 
Grains covered by water icy mantles can play a significant role 
on planetary formation, enabling the formation of planetesimals and the core of gas 
giants protoplanets beyond the \textit{snow line}.  Observations of  the Solar System's 
asteroid belt suggest that  our \textit{snow line} occurred near a disk radius of $\sim$2.7~AU. 
In the outer reaches of our own Solar system, i.e. beyond the snow 
line, most of the satellites and small bodies contain a significant fraction of water ice;
 in the case of comets this fraction is as high as 80\% and the presence of water in the 
upper atmospheres of the four gas giants is thought to be highly influenced by cometary 
impacts such as that of Shoemaker-Levy 9 on Jupiter.  It is possible that it is during the 
later phases of planetary formation that the atmospheres, and indeed the oceans, of the rocky
 planets were formed from water ice contained in the comets and asteroids that bombarded the 
inner Solar system.  Disk models do predict the position of the \textit{snow line} as a function of the 
host star type (e.g. closer to the star for cooler stars).
However, very little is known from 
the observational point of view because the exact location of the \textit{snow line} 
is very hard to infer with near--IR ground-based telescopes.

In the far-IR there is a powerful tool for the detection of water ice and for the 
determination of the 
amorphous/crystalline nature: namely the transverse optical mode at $\sim$44~$\mu$m both from 
crystalline  and amorphous water ice and the longitudinal acoustic mode at $\sim$62~$\mu$m 
arising  only from crystalline  water ice \citep{moor92, malf99}.  
In contrast to the $\sim$3.1~$\mu$m  and $\sim$6.1$~\mu$m water ice bands,
the difference between the amorphous and crystalline phase is very well defined in the far-IR
(see Fig.~\ref{goicoecheaj1_fig-disk-malfait}a) and, again unlike the mid-IR features, the far-IR bands are not 
confused with other solid state features of less abundant species in the ice (\textit{e.g.,} ammonia or methanol).  
In optically thin disks it is 
extremely difficult to use mid-IR absorption to trace water ice and the material is too cold to emit 
in the near- and mid-IR bands.  Hence, these strong far-IR features are robust probes of (1) the presence/absence 
of water ice,  even in  heavily obscured or cold regions without a mid-IR background, and  
(2) the amorphous/crystalline state which provides clues on the formation history of water ice. 
Note that  JWST cannot access these solid-state bands and \textit{Herschel} only has access to the 
$\sim$62~$\mu$m band with much poorer sensitivity and limited bandwidth.

First observed in the far-IR towards the Frosty Leo Nebula \citep{omon90}, water ice has been
 detected in young protoplanetary disks in a few bright sources either in the far-IR using the ISO/LWS 
\citep{dart98, malf99} or mid-IR \citep{pont05}.  The far-IR features were also observed using 
\textit{ISO}/LWS in comets within our own  Solar system \citep{lell98}.  
Since the bands change shape 
\textit{i.e.,}  they narrow for crystalline ice and the peak shifts in wavelength with the temperature, relatively high 
spectral resolution is  needed to extract all the available information from the ice band profiles. 
In its highest  resolution mode, SAFARI provides $R$$\sim$4500 at $\sim$44~$\mu$m  
which is appropriate for  very detailed ice spectroscopy studies.
Indeed SPICA is the only planned mission that will allow water ice to be observed in all 
environments and fully explore its impact on planetary formation and evolution and the emergence 
of \textit{habitable} planets. 

\section{Concluding remarks}

SPICA/SAFARI  will detect far-IR spectral features from key chemical species 
(H$_2$, HD, organic molecules, water vapor) and many bright atomic fine structure lines 
in protoplanetary disks. SAFARI will have enough sensitivity to 
detect them in a volume sufficiently large to make an unbiased study of the chemistry of 
protoplanetary disks out the closest star forming regions ($d\simeq$150\,pc). The detection of the 
gas residual in \textit{transitional} disks at the time planets assemble (at least when gas 
giants and ``hot-Jupiters" form) will provide crucial tests to the planet formation theory itself. 
In the nearest (though more evolved) debris disks  such as Vega or Formalhaut,  SAFARI's large,
fully--sampled FOV and 
unique access to the far-IR water ice features will allow resolving spatially 
the \textit{snow line} (in a single fingerprint), giving a critical insight into the 
role of water ice in the evolution of planetary systems. For more distant disks, only a far-IR space
interferometer (``FIRI"; \textit{e.g.,} Helmich \& Ivison 2009) 
can provide the angular resolution (comparable to ALMA)
needed to resolve spatially the different disk layers where the emission from
the above gas and ice species arises. Although FIRI will be strictly needed in the future
 to fully understand the formation
and evolution of planetary  systems, FIRI is still a  long-term project.
Waiting for FIRI, SAFARI will be able to observe  the critical far-IR range
not covered by JWST and ALMA (at comparable sensitivities) in the next decade, and 
with enough spectral
and angular resolution to provide unique contributions to the study of protoplanetary systems.

\begin{acknowledgements}
We warmly thank the European and Japanese SPICA science teams for fruitful discussions 
and different contributions to develop the ``protoplanetary disks'' case
for SAFARI. 
JRG is supported by a \textit{Ram\'on y Cajal} research contract.

\end{acknowledgements}

\end{document}